# EFFECTS OF THE MASS TRANSFER AND PRESENCE OF THE THIRD COMPONENTS IN CLOSE BINARY STELLAR SYSTEMS

D. E. Tvardovskyi[1], V.I. Marsakova[1,2], I.L. Andronov[2]

[1] Odessa I. I. Mechnikov National University

Odessa, Ukraine, **dtvardovskyi@ukr.net**, **vmarsakova@onu.edu.ua**

[2] Department "Mathematics, Physics and Astronomy", Odessa National Maritime University

Odessa, Ukraine, **tt_ari@ukr.net**

ABSTRACT: In our research, we have studied 6 close binary stellar systems which are eclipsing variables of β Lyrae and W Ursae Majoris types of variability. We have studied their O-C curves. To build them, we used moments of minima, listed in the BRNO database. Also, we used ones, which we obtained as the result of processing of observational data taken from AAVSO database. As the result of the O-C analysis, we detected that all of these stars have parabolic O-C curves, which might be caused by mass transfer from one component to another. In an addition, 3 from researched stars (U Peg, V0523 Cas and WZ Cyg) have superimposed parabolic and cyclic O-C changes that could be caused by presence of the third components in the systems. Also, we calculated minimal possible masses of third components, rates of the mass transfer for these systems and corrected the ephemerides.

## 1. Objects of study

We have selected the variables for the analysis of periods changes using several main criteria. We chose the stars, observable through the small telescope during the most part of the year from Europe. Therefore, the selected stars have to:
- be bright enough (no fainter than 12–14 magnitude in the primary minimum);
- be located in the northern hemisphere of the sky;
- have short period (less than two days);
- have large enough amplitude of variability;
- have significant cyclic or secular period changes.

In an addition, the selected stars should be described at least in one research, in which we can find masses of the system's components (Bell, Adamson & Hilditch, 1987; Borkovits et al., 2005; Giuricin & Mardirossian, 1981; Lee et al., 2011; Kallrath & Strassmeier, 2000; Samec, Faulkner & Williams, 2004).

In this paper, we discuss 6 eclipsing binary stars: BF Aur, SX Aur, U Peg, V0388 Cyg, V0523 Cas and WZ Cyg. All of them are close binary systems, in which two components are deformed by their gravitational interaction.

## 2. Observational data and moments of minima

We used observations from different sources. Most of them were international databases, such as American Association of Variable Stars Observers (AAVSO) [http://www.aavso.org/], Brno Regional Network of Observers (BRNO) [http://var.astro.cz/ocgate/] and Northern Sky Variability Survey (NSVS) (Woźniak et al, 2004). The information about coordinates, constellations, spectral types and others was taken from the "General Catalogue of Variable Stars" (GCVS) (Samus et al., 2017).

For the variable V0523 Cas we also used our own observations obtained in filter R during international Astrocamp "Variable-2017" (http://www.astrokolonica.sk/en/events/variable/).

From the AAVSO database, we used the observations in the filters V, B, R and the visual ones for the variable stars BF Aur, SX Aur, U Peg, V0523 Cas and V0388 Cyg. Observations from the NSVS (in R-band) we used for WZ Cyg.

To collect the long series of minima, we used a large collection of ones for different eclipsing binary stars made by many authors and published in different articles on the web service BRNO [http://var.astro.cz/ocgate/].

Also, we processed all available observations by using the symmetrical polynomial fit (Andrych et al., 2015). That procedure was done for our own observations and those taken from the AAVSO database or from the NSVS, if there was not enough data in AAVSO. Phenomenological modeling of eclipses was reviewed by Andronov, Tkachenko & Chinarova (2017).

## 3. Methodic and hypothesis

### 3.1 O-C analysis

For all available moments of minima, we calculated values of O-C. Then, we built the O−C curves. For the stars with cyclic period changes (which we interpret as periodic within errors of observations), we defined the period and amplitude of the O−C changes.

The period and other parameters have been determined using the non-linear least squares fit, which takes into account simultaneously the algebraic polynomial trend, as well as multi-harmonic periodic wave. For this purpose, we have also used the program MCV (Andronov & Baklanov, 2004), which realizes the algorithm of fitting of the observations with the general formula:

$$x_C(t) = \sum_{k=1}^{p+1} C_k (E-E_0)^{k-1} + \sum_{j=1}^{s} (C_{p+2j}\cos(jw(E-E_0)) + C_{p+2j+1}\sin(jw(E-E_0))), \quad (1)$$

where $x_C$ is the value of function (in our case function is O−C), $E$ is the cycle number, $E_0$ is the fixed value of $E$, which we set as initial, $w = 2\pi/T$ is the rotation frequency, $T$ is the period of changes of orbital period ($P$) of binary system; $j, k, p, s$ are the integer numbers. The first part describes an algebraic polynomial trend of degree $p$ and the second one – a trigonometric polynomial of power $s$. The coefficients $C_i$, $i = 1 \ldots (p+1+2s)$, which may be determined using the least squares method (Andronov, 2003). The O−C curves and their approximations are shown at Fig. 1. The coefficients of the polynomial trend were used for the correction of the ephemerids.

In the case of parabolic trend ($p = 2$), two coefficients ($C_1$ and $C_2$) mean the same as in the previous case. The coefficient ($C_3$) defines the rate of the secular period changes (Andronov, 1991, 2003).

Corrected ephemerids are listed in Table 1.

### 3.2 The rate of the mass transfer

If the star has a parabolic O−C curve or a cyclic one with parabolic trend, its period changes steadily. As the result of the O−C analysis, we detected that six of the researched stellar systems have secular period changes, which we interpreted as the mass transfer from one of the components to another. To calculate the rate of the mass transfer, we used the formula (Huang, 1963).

$$\dot{M} = \frac{1}{3}\frac{\dot{P}}{P}\frac{M_1 M_2}{(M_1 - M_2)} = \frac{1}{3}\frac{P'}{P^2}\frac{M_1 M_2}{(M_1 - M_2)}, \quad (2.2)$$

In this formula: $\dot{M} = \dot{M}_1$ is the rate of the mass transfer (measured in solar masses per day), $M_1, M_2$ are masses of the components in close binary system, $P$ is the period of variability, $\dot{P}$ is the derivative of the period with time, $P' = 2C_3$ is the derivative of the period with cycle number (e.g. Andronov, 1991).

### 3.3 The third components hypothesis

For three systems, we detected cyclical period changes as the result of the analysis of the O−C curve. We interpret as periodic within errors of observations and supposed that they might be caused by the presence of a hypothetical third component (star or planet) in each of these systems. In our hypothesis, this component, because of its gravity, makes the binary system rotate around the common barycenter and causes the well-known light-time effect (Wolf, 2014).

In our model, we suppose that the distance between two components of binary system is much less than distance from binary system to the third component. So we can take the binary system as a single object with mass $M_1 + M_2$. Therefore, we can describe the motion of the binary system and the third body with these formulas:

$$A^3 = \frac{GMT^2}{4\pi^2} = \frac{G(M_1 + M_2 + M_3)}{\omega^2} \quad (2.3)$$

$$a_{12} = c\tau = A\frac{M_3}{M_1 + M_2 + M_3}\sin i \quad (2.4)$$

Formula (2.3) is the third Kepler's law; formula (2.4) comes from barycenter position (Tatum, 2017). In these formulas $M_1, M_2, M_3$ are masses of the components, $i$ is angle of the orbit's inclination, $A$ is semi major axis of the third component relatively to the binary system, $a_{12}$ is semi major axis of the binary system relatively to the common barycenter, $\tau$ is the semi-amplitude of the O−C curve, $T$ is the orbital period of the third component, $\omega$ is the angular velocity that corresponds $T$. There are three constants: $G$ is the gravitational constant; $c$ is the speed of light.

So we obtain:

$$\sin i = \frac{c\tau(M_1 + M_2 + M_3)}{AM_3} \quad (2.5)$$

We do not know neither $\sin i$, nor $M_3$. Therefore, we tabulated mass with a small step and then computed the sine of the inclination angle. Since sine of any angle cannot be more than unity, it gives us the limit of minimal mass of the third component. Thus, we built graphs, which show us relations between the mass of the third component and the orbit inclination (Fig. 2). Minimal masses of possible third components are listed in Table 2.

In addition, we calculated the errors of the minimal mass of the third body and the rate of the mass transfer by using formulas below. To obtain them, we used standard methods, which are described (Korn & Korn, 2000).

Formula of the error of the mass transfer:

$$\sigma = \sqrt{\left(2\frac{\sigma P}{P}\right)^2 + \left(\frac{\sigma P'}{P'}\right)^2 + \left(\frac{\sigma M_1}{M_1} - \frac{\sigma M_1}{M_1 - M_2}\right)^2 + \left(\frac{\sigma M_2}{M_2} + \frac{\sigma M_2}{M_1 - M_2}\right)^2}$$

where $\sigma$ is the statistical error of the mass transfer, $\Delta$ is the shift and similarly for other parameters.

To obtain the errors for minimal mass of the third component we use the similar approach.

Formula of the minimal mass error of the third component may be expressed as:

$$\sigma M_3 = \frac{M_3}{\left(\frac{1}{M_3} - \frac{2}{3M}\right)}\sqrt{\left(\frac{\sigma \tau}{\tau}\right)^2 + \frac{4}{9}\left[\left(\frac{\sigma M_1}{M_1}\right)^2 + \left(\frac{\sigma M_2}{M_2}\right)^2 + \left(\frac{\sigma T}{T}\right)^2\right]}$$

$\Delta$ and $\sigma$ mean the same as in formula (2.6).

### 4. Conclusions

Our main results are given in Tables 1 and 2.

The rates of the mass transfer in the studied systems SX Aur, and U Peg are in a good agreement with results of other authors (Bell et al., 1987), (Borkovits et al., 2005). For WZ Cyg, we obtained the value, which is three times smaller than in (Lee et al, 2011). This can be caused by low accuracy and little quantity of data taken by us and other authors from older researches. For BF Aur and V0388 Cyg, we calculated the mass transfer rates for the first time.

For 3 systems (SX Aur, U Peg and V0523 Cas), the third components were suspected by other authors (Borkovits et al., 2005; Lee et al., 2011; Samec, Faulkner & Williams, 2004) but for U Peg the masses of the third component haven't been estimated earlier by other authors. For WZ Cyg, our results concerning periods of period changes are in good agreement with (Lee et al., 2011) (within our errors). Mass of the third body were estimated for WZ Cyg (Lee et al., 2011) it is comparable in order of values with our estimates. For V0523 Cas our estimation of the mass of the third body is approximately 20 percent smaller than one in (Samec, Faulkner & Williams, 2004).

As we can see from the Table 2, the majority of the third components have masses like stars. This fact may be caused by low accuracy of eclipse moments, especially moments that calculated decades ago (we analyzed the data obtained during about 100 years). This low accuracy does not allow to detect small period changes due to planet gravitation. But the usage of the international databases of observations and minima of eclipsing variable stars is very helpful for the analysis of the long-term period changes of these objects.

The alternative hypothesis for the explanation of cyclic period changes are discussed in (Zhu et al., 2012) and (Borkovits et al., 2005). They are magnetic activity of the second component of binary system or, in case stable changes and cyclic variations, cyclic changes in mass transfer. Confirmation or refutation of our hypothesis requires continuous spectroscopic observations of these systems.

We made similar calculations for other three systems (Tvardovskyi & Marsakova, 2015).

**Acknowledgments.** We are grateful to amateur astronomers for their observations published in the AAVSO and BRNO international databases that made this research possible. This work is a part of the international campaigns "Inter-Longitude Astronomy" (Andronov et al., 2017) and "Astroinformatics" (Vavilova et al., 2017).

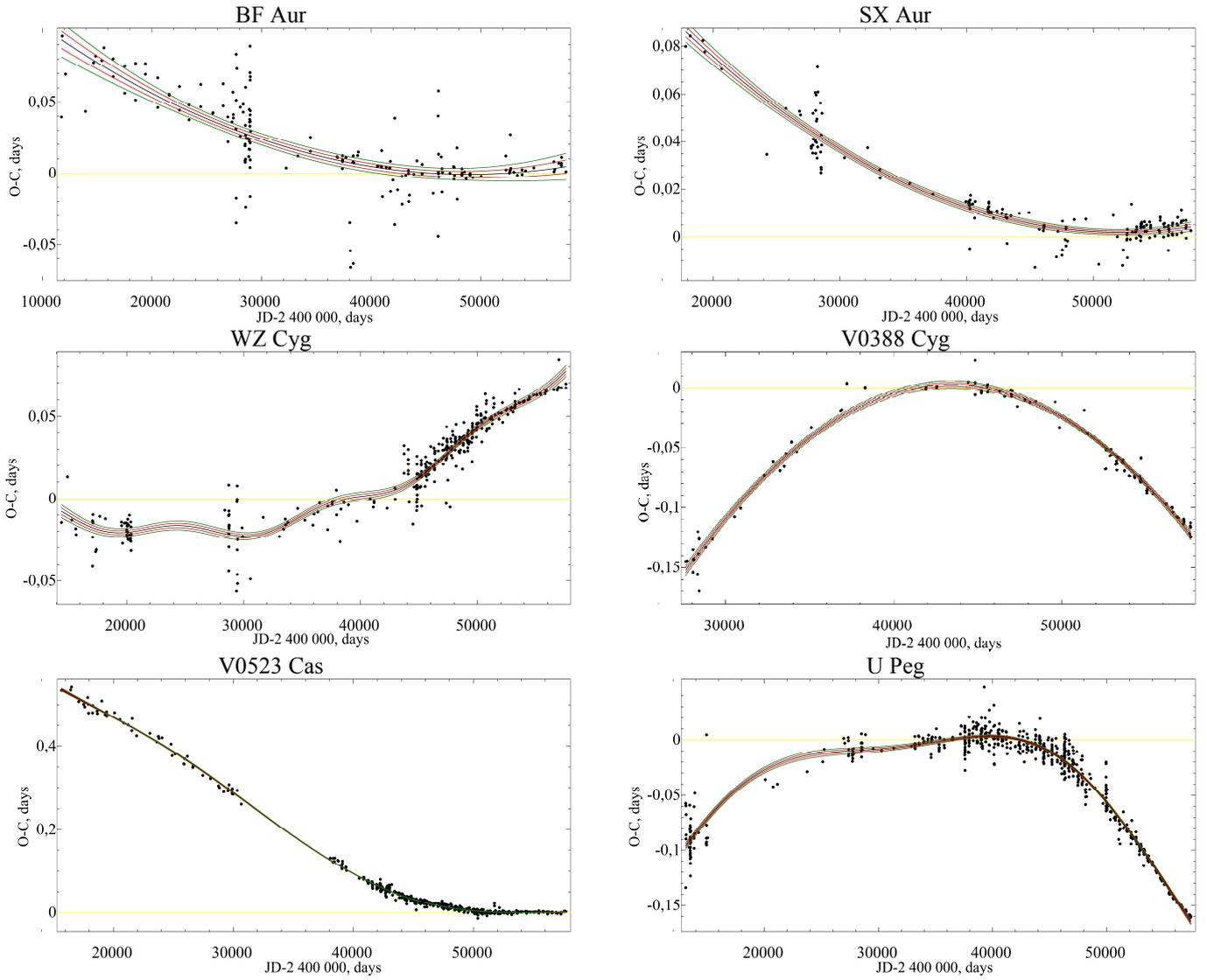

Fig. 1. O−C curves for the studied systems. On the horizontal axis we set Julian date without 2400000; on the vertical one – value of O-C. Designation of each system indicated near corresponding curve. Solid lines show approximations, which take into account simultaneously the algebraic polynomial trend, as well as multi-harmonic periodic wave (using the MCV program (Andronov & Baklanov, 2004)). Red and green lines correspond to confidence interval of one and two root-mean-square deviations.

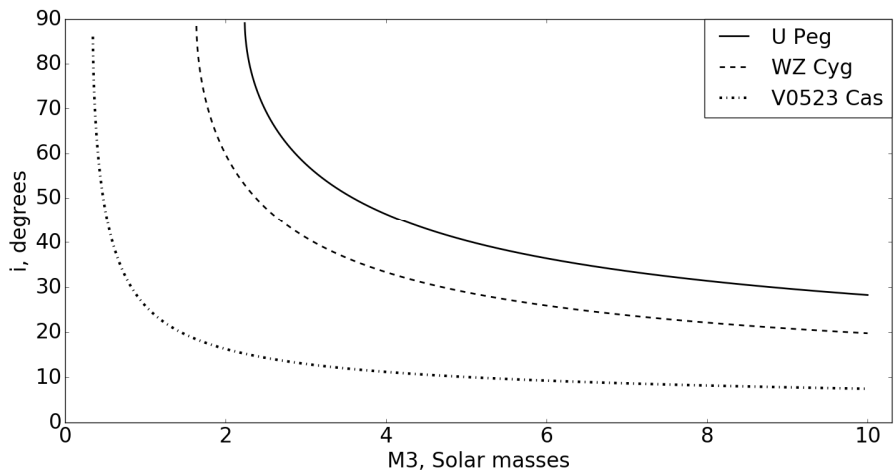

Fig. 2. Dependence of orbital inclination of the third component on its mass (measured in solar masses) for the systems U Peg, WZ Cyg and V0523 Cas.

Table 1. Ephemerides before and after correction. Here $E$ is the number of cycle. BJD means that the value was corrected for the Earth motion around the Sun (barycentric corrections).

| Stellar system | Ephemerid before correction (GVCS [13]) Min I BJD – 2400000 | Ephemerid after correction Min I BJD - 2400000 |
|---|---|---|
| V0388 Cyg | 41953.3373+0.8590372· $E$ | (41953.337±0.003)+(0.8590372± 2·10$^{-7}$) · $E$ −(48±2)·10$^{-11}$· $E^2$ |
| BF Aur | 52500.950+1.5832232· $E$ | (52500.947±0.002)+(1.5832249 ± 2·10$^{-7}$) · $E$+(17±3)·10$^{-11}$· $E^2$ |
| SX Aur | 52500.3179+1.2100855· $E$ | (52500.3119±0.0008)+( 1.2100872 ± 7·10$^{-8}$) · $E$+ (105±8)·10$^{-12}$· $E^2$ |
| WZ Cyg | 40825.475+0.5844659· $E$ | (40825.6081±0.0007)+(0.58446590±7·10$^{-8}$) · $E$+(1±0.4)·10$^{-10}$· $E^2$ |
| U Peg | 36511.66823+0.37478143· $E$ | (47070.5200±0.0005)+(0.37477680±3·10$^{-8}$) · $E$−(4.8±0.2)·10$^{-10}$· $E^2$ |
| V0523 Cas | 57330.111+0.2336933· $E$ | 57330.148 + 0.2337238 · $E$ + (7.66±0.08)·10$^{-9}$ |

Table 2. Calculated characteristics of the period changes, minimal masses of the third components and the rates of the mass transfer for researched systems.

| The name of the stellar system | The mass of the binary system | Period of the period changes | Semi-amplitude of the cyclic period changes | Minimal mass of the third body | The rate of the mass transfer |
|---|---|---|---|---|---|
| | Solar masses | days | Days | Solar masses | Solar masses per year |
| SX Aur | 16 (Bell, Adamson & Hilditch, 1987) | — | — | — | (3.1±0.3)·10$^{-7}$ |
| U Peg | 1.5 (Borkovits et al., 2005) | 22633±1114 | 0.084±0.003 | 0.32±0.13 | (−3.7±0.3)·10$^{-8}$ |
| V0388 Cyg | 4.5 (Giuricin & Mardirossian, 1981) | — | — | — | (−2.7±0.3)·10$^{-7}$ |
| WZ Cyg | 2.6 (Lee et al., 2011) | 17196±1045 | 0.0047±0.0009 | 0.123±0.06 | (2.0±0.3)·10$^{-7}$ |
| BF Aur | 10.1 (Kallrath & Strassmeier 2000) | — | — | — | (−1.7±1.5)·10$^{-6}$ |
| V0523 Cas | 1.18 (Samec, Faulkner & Williams, 2004) | 33976±867 | 0.03071±0.00057 | 0.343±0.006 | (7.6±0.8)·10$^{-8}$ |